\documentclass[12pt]{article}
\usepackage{mathtext}
\usepackage{graphicx}
\usepackage{amsfonts}
\usepackage{amssymb}
\usepackage{amsmath}
\numberwithin{equation}{section}

\renewcommand{\d}{\delta}

\newcommand{\w}{\omega}
\newcommand{\s}{\sigma}
\newcommand{\la}{\lambda}

\begin{document}
\title{Collective micro-causality in homogeneous quantum ensembles}

\author{Y.I.Ozhigov\thanks{ozhigov@cs.msu.su}\\
Moscow State University of M.V.Lomonosov, \\
VMK Faculty,\\
Institute of Physics and Technology RAS}
\maketitle

\begin{abstract}
It is shown that for quantum ensembles consisting of equal particles, the collective micro-causality does not contradict the quantum theory. The amplitudes of the states of such an ensemble can be divided into small portions, for each of which evolution in time will be accurately determined, without any randomness. For ensembles of dissimilar particles, whose Hamiltonian is represented as a sum of operators belonging to homogeneous systems, micro-causality will take place on separate small segments, on which only the selected component will act. In this case, the duration of the terms for different ensembles will be different. Micro causality is a consequence of the quantization of amplitudes, which plays a peculiar role for complex systems in which the amplitudes of individual states are very small. For such systems, the stochastic description, typical in the standard quantum theory should be replaced by a causal, i.e. deterministic one. In the article micro-causality is illustrated by the example of ensembles of two-level atoms in an optical cavity, considered under the Tavis-Cummings model in the rotating wave approximation; however, the conclusions are general.
\end{abstract}

\section{Introduction}

"God does not play dice" - Einstein's phrase expresses faith in reasonableness, that is, the determinism of the world. It is believed that quantum theory refutes this reasonability, and we are doomed to see the world only through the prism of stochastic processes in which every event has no cause and occurs purely randomly (\cite{BE}, \cite{Ge}). And this is true, but only for "simple" processes, the description of which does not require the involvement of the notion of entanglement and therefore can be given - approximately - in the framework of classical mechanics;  the successes of quantum theory in the 20-th century are connected with such processes. The peculiarity of them is the possibility to consider the environment as Markovian, that is without long-term memory, within the framework of an open quantum system (\cite{BP}) - a concept that works very well for almost all systems, except for living beings, to which it is not applicable in principle.

The extension of quantum theory to complex multi-partial processes, which no longer have a quasi-classical description, requires the involvement of the Hilbert space formalism from the very beginning, and therefore for such processes the universal model is the Feynman quantum computer (\cite{Fe3}). The understanding of the work of this, while hypothetical, device radically at odds with the intuitive perception of the principle of causality, that is, with determinism.

It is impossible to express the work of a quantum computer with the help of formulas, as is done for model problems of quantum physics (see, for example, \cite{Ak}), since this contradicts the existence of fast quantum algorithms (see, for example, \cite{Gr}), the results of which is fundamentally impossible to predict faster than they will arise in reality; this fact makes quantum computer a very special device. To simulate its work, for example, for debugging its gates, we need to use existing computers that work on the principle of determinism.

This determinism manifests itself in the fact that in practical numerical calculations the amplitude of quantum states is never infinitely small: in machine calculations there is always a virtual "amplitude quantum" $\epsilon>0$, such that if the absolute value of some amplitude becomes less than $\epsilon$, it is considered to be zero. This, of course, makes it impossible to model a scalable quantum computer. For it to work, you need to receive the states $|\Psi\rangle=\sum\limits_{j\in J}\la_j|j\rangle$ with nonzero $\la_j$, such that the power $|J| $ of the set $J$ grows exponentially and storage in the computer memory of all its elements is absolutely impossible. The presence of an "amplitude quantum" is an equivalent form of such a prohibition, since the decomposition $|\Psi\rangle$ must contain coefficients $\la_j$ such that $|\la_j|\leq |J|^{-1/2}$.

Since there is no alternative to machine calculation of quantum computer constructions, we must consider the possibility that the "amplitude quantum" has a physical meaning. Namely, in ensembles of $n$ particles, for which $|J|=exp(n)$, there is a natural decoherence associated with the disappearance of the state components $|j\rangle$ with too small amplitude $\la_j$. This source of coherence loss, in the case of the approximate equality of the absolute values of all $|\la_j|$, would lead to the elimination of coherence as such, and therefore to real determinism.

Finding ways to such a deterministic description of complex systems is a profound goal of the theory of quantum computer.

However, determinism must somehow be manifested in the standard quantum theory, where $\epsilon$ can be directed to zero, that is, the amplitude quantum must consistently fit into the analytical technique applicable to experiments. It turns out that quantum theory allows for a paradoxical way of introducing causality - in the form of quantization of the amplitudes of elementary events, in such a way that each individual portion of the amplitude under certain conditions, will have causality. Such causality occurs for ensembles consisting of equal particles, which is typical for the prototypes of a quantum computer, for example, ensembles of two-level atoms interacting equally with a field inside an optical resonator. For dissimilar ensembles, the amplitude quantization must be performed separately for different parts of the system under consideration, so that different parts of the dissimilar ensemble will evolve at different rates. In this paper, this general thesis is illustrated by an example of ensembles of two-level atoms in an optical cavity within the framework of the Tavis-Cummings model with RWA approximation.

The standard matrix quantum formalism is based on the principle pluralism, which consists in the reproduction of each portion of the amplitude, in which its imaginary "trajectory" becomes branching - the essence of matrix multiplication, which serves as a formal representation of the evolution of the quantum state in time\footnote{Expressive illustration of this principle is given in the book \cite{Fe1}.}. However, it turns out that under certain conditions it is possible to assign the special path to each small portion of the amplitude, in which there will be no branches: this path is determined exactly. This condition: real particles included in the ensemble must be identical in relation to the Hamiltonian, that is to play in the evolution absolutly the same role.

The limiting case of a homogeneous ensemble is an ensemble consisting of one particle. The quantum dynamics of the density of one particle, in the case of the usual dynamic Hamiltonian, will be described by a swarm of virtual particles, which is subject to the Hamilton-Jacobi equations with a  quantum pseudo-potential of Bohm (\cite{B}). This approach is called quantum hydrodynamics\footnote{Quantum hydrodynamics is aimed at complete reformulation of quantum mechanics as such, for example, it pretends to explain the presence of spin in particles. In contrast, quantization of amplitudes is necessary to find "metagenes" as causes of complex phenomena. }, and it is generalized to ensembles of particles moving under the law of quantum mechanics with Hamiltonian obtained by the standard procedure from classical expressions for energy.

Quantum hydrodynamics of de Broglie - Bohm is applicable only to such ''quasi-classical'' Hamiltonian and only in the form of quantization of the probability of occurrence of a particle in a given space-time domain. The micro-causality in quantum hydrodynamics is thus restricted to only such Hamiltonians, and the phase of the wave function can be calculated after averaging over the impulses in each elementary space-time cell; this can be done only if the computational resource is concentrated on one particle. In Hamiltonians of Tavis-Cummings type dynamic characteristics of the individual atoms do not exist at all, and the swarm of de Broglie - Bohm is not applicable to establish the properties of causality in TC - like models, as well as in models of quantum computers.

Amplitudes in the form of individual quanta can be directly manifested in complex systems, with a large number of particles, so that the probability is "smeared" by a huge number of mutually orthogonal basic states\footnote{Their orthogonality is guaranteed by the choice of grain for the resolution of space and time, which must be done beforehand, because even charges of particles depend on this grain.}. In this case, the micro-causality will mean that the trajectory of the whole system is uniquely determined by its current configuration, and the probabilistic nature of quantum dynamics becomes a strict determinism. The search for such determinism is associated with interpretations of quantum theory, in particular, with the so-called contextual dependence of the experimental results on the environment (see \cite{Kh}). Important here is the exact wording, because the contextual dependence is often erroneously interpreted as the introduction of certain "hidden variables" of quantum theory. In fact, such dependence is simply taking into account the quantum nonlocality of entangled states confirmed in numerous experiments (see, for example, \cite{Aspe},\cite{Zei}).

We show, however, that micro-causality holds also for closed systems, for which their individual components have the same type of behavior, and which can be well described by standard algebraic means.

\section{Tavis-Cummings model}

The Jaynes-Tavis-Cummings- (TC) model describes the dynamics of a group of two-level atoms in an optical cavity interacting with a single-mode field inside it. The value of this approach is that it allows to describe very complex interaction of light and matter in the framework of finite - dimensional computational model that in addition has the physical prototypes, among which the most developed is an optical cavity - resonator and a group of two level atoms held in it by optical tweezers.

Tavis-Cummings Hamiltonian (see \cite{D},\cite{JC},\cite{Tav}) $H_{TC}$ for $n$ two-level atoms in an optical cavity whose photon frequency coincides with that of an atomic excitation has the form:
\begin{equation}
H_{TC}=H_c+H_a+H_{int},\ H_c=\hbar\omega a^+a,\ H_a=\hbar\omega\sum\limits_{j=1}^n\s_j^+\s_j,\ \ H_{int}=
\sum\limits_{j=1}^ng_j(a^++a)(\s_j^++\s_j),
\label{TC}
\end{equation}
where $a^+,a$ are mutually conjugated field operators of creation and annihilation of a photon, $\sigma^+_j,\sigma_j$- raising and lowering mutually conjugated operators for $j$-th atom, acting on its ground ($|0\rangle_j$) and excited ($|1\rangle_j$) states as $\s_j |0\rangle_j=0,\ \s_j |1\rangle_j=|0\rangle_j$ (by default, it assumes that the rest of the state components are affected by the identity operator).  Here $g_j$ is the force of interaction between $j$-th atom and the field. 

We shall consider the case of weak interaction where $g_j\ll \hbar\w$. Here the rotating wave approximation (RWA) is applicable that is we can omit not conserving energy summands $a^+\s_j^+$ and $a\s_j$, and write the energy of field-atom interaction in the simpler form $H_{int}^{RWA}=a\sum\limits_jg_j\s_j^++a^+\sum\limits_jg_j\s_j$.

\section{Connected states}

Let $H$ be a Hamiltonian in the state space of $n$ qubits, each of which is associated with a real two-level particle. In particular, it can be TC Hamiltonian. Let $S_n$ be a group of permutations of qubits that are naturally extended to operators on the whole space of quantum states ${\cal H}$, namely: on the basic states the permutation $\eta\in S_n$ acts directly, and $\eta\sum\limits_j|j\rangle=\sum\limits_j\eta|j\rangle$.

We denote by $G_H$ the subgroup in $S_n$, consisting from all permutations of qubits $\tau$, such that $[H,\tau]=0$. Let $A\subseteq\{ 0,1,...,2^n-1\}$ be a subset of basic states of $n$- qubit system. Its linear envelope $L(A)$ we call connected with respect to $H$, if for any two states $|i\rangle,\ |j\rangle\in A$ there exists the permutation of qubits $\tau\in G_H$, such that $\tau (i)=j$. A state $|\Psi\rangle$ of $n$- qubit system we call connected with respect to $H$, if it belongs to subspace connected with respect  to $H$ and $H|\Psi\rangle\neq 0$. 

The connectivity of the state means that all its nonzero components are obtained by one from the other by permutations of those particles that are equivalent with respect to a given Hamiltonian, that is, behave in relation to it in the same way. 

\bigskip
{\it Example 1}.
Let $Q_1,Q_2,...,Q_s$ be a partition of the set f all qubits to subsets, such that for all $j=1,2,...,s$ any transposition $\tau$ of two qubits from the set $Q_j$ commutes with the Hamiltonian $H$: $[H,\tau]=0$. We consider the state $|\Psi\rangle=\sum\limits_j\la_j|j\rangle$. If for all basic states $|j_1\rangle$ and $|j_2\rangle$, such that $\la_{j_1}$ and $\la_{j_2}$ are nonzero and any $r\in \{ 1,2,...,s\}$ the strings of values of qubits from $Q_r$ in the states $j_1$ and  $j_2$ have the same Hamming weight, then the state $|\Psi\rangle$ is connected with respect to $H$. 
\bigskip

{\it Proposition.

If $|\Psi\rangle=\sum\limits_j\la_j|j\rangle$ is connected with respect to $H$, then any two columns of the mathrix $H$ with numbers $j_1,\ j_2$, such that  $\la_{j_1}$ and $\la_{j_2}$ are nonzero, differ one from the other only by the permutation of elements.}
\bigskip

Indeed, by the definition of connectivity, for such basic states $j_1$ and $j_2$ there exists $\tau\in G_H$, such that $j_2=\tau(j_1)$. Columns with numbers $j_1,\ j_2$ consist of amplitudes of states  $H|j_1\rangle$ and $H|j_2\rangle$ correspondingly. From the condition of commutation we have $\tau H|j_1\rangle=H\tau |j_1\rangle=H|j_2\rangle$, and this means that the column $j_2$ is obtained from the column $j_1$ by the permutation of elements induced by $\tau$. Proposition is proved.

\bigskip
{\it Example 2}. We consider as $H$ Tavis-Cummings Hamilton $H_{TC}^{RWA}$ with zero detuning of frequencies for $n$ atoms, which have the equal force of interaction with the field. Then $G_H=S_n$, that can be checked straightforwardly: for any transposition $\tau=(i,j)$ of two atomic qubits and basic state $|m\rangle$ the coincidence of states $\tau H|m\rangle$ and $H\tau |m\rangle$ follows from the equality between the forces of interaction with the field for all atoms. This also gives that each permutation of atomic qubits commutes with Hamiltonian. Let $\tilde{\cal H}^n_{k_a,k_p}$ be the linear envelope of such basic states, in which atomic parts have energy $k_a\hbar\omega$ (contain $k_a$ units), and the photon part is $|k_p\rangle_{ph}$, where $k_a,k_p$ are natural numbers. Then $\tilde{\cal H}^n_{k_a,k_p}$ will be connected with respect to $H_{TC}^{RWA}$. 
\bigskip

\section{Event quantization}

Our goal is to show that if the state of $ |\Psi\rangle$ is connected with respect to the Hamiltonian $H$, then the amplitudes of all the basic states in $ |\Psi \rangle$ can be divided into small portions - quanta of amplitude, so that for each quantum its trajectory under the action of Hamiltonian $H$ on a small interval of time will be uniquely determined, in particular, will be uniquely determined also with what other quantum it will cancel in the calculation of the amplitudes for the subsequent state in unitary evolution $exp(-iHt/\hbar)$.  

Let $|\Psi\rangle$ be arbitrary state of $n$ qubit system, connected with respect to $H$, which expansion on basic states has the form $|\Psi\rangle=\sum\limits_j\la_j|j\rangle$. 

We introduce the important notion of amplitude quantum as a simple formalization of the transformation of a small portion of the amplitude in evolution on a small time interval during the transition between different basic states. Let $T=\{ +1,-1,+i,-i\}$ be the set of 4 elements, which are called types of amplitude: positive real, negative real, and the same imaginary. The product of types is defined in a natural way: as a product of numbers. The amplitude quantum of $\varepsilon > 0$ is the list of the form
\begin{equation}
\label{quanta}
\kappa=(\varepsilon,id, |b_{in}\rangle, |b_{fin}\rangle, t_{in},t_{fin})
\end{equation}
where $|b_{in}\rangle,\ |b_{fin}\rangle$ are two different basic states of the system ''atoms+photons'', $id$ is a unique identification number that identifies this quantum among all others, $t_{in},t_{fin}\in T$.  The transition of the form  $|b_{in}\rangle\rightarrow\ |b_{fin}\rangle$ we call state transition, $t_{in}\rightarrow t_{fin}$ type transition.  We select the identification numbers so that when they match all the other attributes of the quantum also coincide, that is, the identification number uniquely determines the amplitude quantum. In this case, there must be an infinite number of quanta with any set of attributes, with the exception of the identification number. Thus, we will identify the amplitude quantum with its identification number, without specifying it in the future. We introduce notations:
$$
t_{in}(\kappa)=t_{in},\ t_{fin}(\kappa)=t_{fin},\ s_{in}(\kappa)=b_{in},\  s_{fin}(\kappa)=b_{fin}.
$$

State transitions and types of amplitude quanta point in fact, how this state should change in time, and their choice depends on the Hamiltonian; the size of the amplitude quanta indicates the accuracy of the discrete approximation of the action of the Hamiltonian with the help of them.

A set $ \theta $ of amplitude quanta of the size $ \varepsilon$ is called a quantization of the amplitude of this size if the following condition is satisfied:

{\bf Q}. There is no such quanta $\kappa_1$ and $\kappa_2$ in the set $\theta$ that their state transition is the same, $t_{in}(\kappa_1)=t_{in}(\kappa_2)$ and $t_{fin}(\kappa_1)=-t_{fin}(\kappa_2)$; and there is no such quanta $\kappa_1$ and $\kappa_2$, that their state transition is the same and $t_{in}(\kappa_1)=-t_{in}(\kappa_2)$ . 

\bigskip

The condition {\bf Q} means that at the transition described by the symbol "$\rightarrow$" the resulting value of any amplitude quantum cannot cancel with the resulting value of the other - simular quantum. 

We introduce the notation $\theta (j)=\{ \kappa:\ s_{in}(\kappa)=j\}$. If $|j\rangle,\ |i\rangle$ are basic states, $t_i,t_j\in T$ are the types, $\theta$ is amplitude quantization, we introduce the notation $K_\theta (i,j,t_i,t_j)=\{ \kappa\in\theta(j),t_{in}(\kappa)=t_j,t_{fin}(\kappa)=t_{i}, s_{fin}(\kappa)=i\}$.

For any complex $z$, we define its relation to the type $t\in T$ naturally: $[z]_t=|Re(z)|$, if $t=+1$ and $Re(z)>0$, or  $t=-1$ and $Re(z)<0$;  $[z]_t=|Im(z)|$, if $t=+i$ and $Im(z)>0$, or $t=-i$ and $Im(z)<0$; $[z]_t=0$ in all other cases. 

We call $\theta$- shift of a state $|\Psi\rangle$ the state $|\theta \Psi\rangle=\sum\limits_i\mu_i|i\rangle$, where for any basic $|i\rangle$
 \begin{equation}
\label{shift}
\mu_i=\langle i|\theta\Psi\rangle=\varepsilon\sum\limits_{\kappa\in\theta:\ s_{fin}(\kappa)=i}t_{fin}(\kappa).
\end{equation}
Amplitude quantization $\theta$ factually determins the transition $|\Psi\rangle\rightarrow |\theta \Psi\rangle$.

We fix the dimension of $dim ({\cal H})$ of the state space, and we estimate (from above) the considered positive quantities: the time and the size of the amplitude quantum up to the order of magnitude, counting all constants depending only on the independent constants: $dim ({\cal H})$ and on the minimum and maximum absolute values of the elements of the Hamiltonian $h$. In this case, the term strict order will mean the upper and lower evaluation by positive numbers, depending only on the independent constants.

We show that for a state $|\Psi\rangle$ connected with respect to $H$ and arbitrarily small $\varepsilon>0$, there exists $\delta>0 $ of strict order $\varepsilon$ and a quantization of the amplitude $ \theta $, which size is of the strict order of $\varepsilon^2$, such that $\theta$ approximates with an error of order $\varepsilon $ the state $ |\Psi \rangle$, and the state $ \delta H |\Psi\rangle$ with the same error is approximated by $\theta$- shift  $|\theta \Psi\rangle$.

\bigskip

{\it Theorem. 

Let $|\Psi\rangle$ be a state connected with respect to $H$. 
Then for any number $\varepsilon>0$ there exists amplitude quantization $\theta$ of the size $\epsilon$ of the order $\varepsilon^2$, the number $\varepsilon_1$ of the order $\varepsilon$ and the number $c$ of the strict order $1$, such that the following two conditions are satisfied.

1) For any basic state $j$ 
\begin{equation}
|\epsilon (\sum\limits_{\kappa\in R_+}1-\sum\limits_{\kappa\in R_-}1+i(\sum\limits_{\kappa\in I_+}1-\sum\limits_{\kappa\in I_-}1))-\langle j|\Psi\rangle|\leq\varepsilon
\label{amplitude_expansion}
\end{equation}
where $R_+=\{ \kappa:\ \kappa\in\theta (j),t_{in}(\kappa)=+1\}$, $R_-=\{ \kappa:\ \kappa\in\theta (j),t_{in}(\kappa)=-1\}$, $I_+=\{ \kappa:\ \kappa\in\theta (j),t_{in}(\kappa)=+i\}$, $I_-=\{\kappa:\ \kappa\in\theta (j),t_{in}(\kappa)=-i\}$ and

2) For any basic statesa $|j\rangle,\ |i\rangle$ and any types $t_{j}, t_{i}\in T$ the following inequality takes place 
\begin{equation}
\label{passage_expansion}
|\epsilon\left(\sum\limits_{\kappa\in K_\theta (i,j,t_i,t_j)}1\right)-c[\langle j|\Psi\rangle\langle i|H|j\rangle]_{t_j}|\leq \varepsilon_1.
\end{equation}
}

{\it Proof. } The meaning of item 1) is that the amplitude quantization gives a good approximation of amplitudes of the state $|\Psi\rangle$; the meaning of item 2) is that the $\theta$-shift corresponding to this quantization $\theta$ gives an approximation of the state $сH|\Psi\rangle$ with error of the order $\varepsilon$   (see Corollary below).

Let we be given a state $|\Psi\rangle=\sum\limits_j\la_j|j\rangle$ connected with respect to $H$  and a number $\varepsilon>0$. For $|j\rangle$ with nonzero $\la_j\neq 0$ we represent the amplitude as
\begin{equation}
\la_j=\langle j|\Psi\rangle\approx sign_{re}( \underbrace{\varepsilon+\varepsilon+\ldots +\varepsilon}_{M_j})+sign_{im}i( \underbrace{\varepsilon+\varepsilon+\ldots +\varepsilon}_{N_j} )
\label{quanta_expansion}
\end{equation}
where $sign_{re}\varepsilon M_j+sign_{im}i\varepsilon N_j\approx \la_j$ is the best approximation of the amplitude $\la_j$ within $\varepsilon$; $M_j,\ N_j$ are natural numbers. The item 1) of the Theorem will be then almost satisfied, only without determining of the final states $|i\rangle$ and final types $t_i$, which depend on the Hamiltonian. 

We approximate each element of the Hamiltonian in the same way we approximate the amplitudes of the initial state:
\begin{equation}
\label{hamapp}
\langle i|H|j\rangle\approx \pm(\underbrace{\varepsilon+\varepsilon+...+\varepsilon}_{R_{i,j}})\pm i (\underbrace{\varepsilon+\varepsilon+...+\varepsilon}_{I_{i,j}})
\end{equation}
where $R_{i,j},\ I_{i,j}$ are natural numbers; real and imaginary parts-with an accuracy of $\varepsilon$ each, and signs in front of the real and imaginary parts are chosen on the basis that this approximation should be as accurate as possible for the selected $\varepsilon$.

The amplitude of the resulting state $H|\Psi\rangle$ are obtained by multiplying the various expressions \eqref{quanta_expansion} on all sorts of expression \eqref{hamapp}:

\begin{equation}
\label{mult}
\la_j\langle i|H|j\rangle\approx (sign_{re}M_{j}\varepsilon+i\ sign_{im}N_{j}\varepsilon )(\pm R_{i,j}\varepsilon\pm i\ I_{i,j}\varepsilon).
\end{equation}

Each occurrence of the expression $\varepsilon^2$ in the amplitude of the resulting state after opening the brackets in the right side of \eqref{mult} is obtained by multiplying the specific occurrence of $\varepsilon$ to the right side of \eqref{quanta_expansion} for a certain occurrence of $\varepsilon$ to the right side of \eqref{hamapp}. The problem is that the same occurrence of $\varepsilon$ in \eqref{quanta_expansion} corresponds not to one but to several occurrences of $\varepsilon^2$ in the result, and therefore we cannot map amplitude quanta directly to occurrences of $\varepsilon$ in \eqref{quanta_expansion}. 

How many occurrences of $\varepsilon^2$ in the amplitude of the state $H|\Psi\rangle$ corresponds to one occurrence of $\varepsilon$ in the approximation of the amplitude $\la_j=\langle j|\Psi\rangle$ the state $|\Psi\rangle$? This is the multiplicity of the given occurrence of $\varepsilon$ - equal to $\sum\limits_i (R_{i,j}+I_{i,j}).$ These numbers can be different for any Hamiltonian $H$ and state $|\Psi\rangle$. However, since $ |\Psi\rangle$ is connected with respect to $H$, by virtue of the Proposition, the columns of the matrix with different numbers $j$ for nonzero $\la_j$ will differ only by permutation of the elements, so the numbers $\sum\limits_i(R_{i,j}+I_{i, j})$ for different $j$ will be the same.

We introduce the notation $\nu=\sum\limits_i(R_{i,j}+I_{i,j})$ is the number of occurrences of $\varepsilon$ in any column of the decomposition matrix \eqref{hamapp}. By virtue of definition of the connectivity for any $j=0,1,2,..., N-1$ such that $\la_j\neq 0$ one of the numbers $\langle i|H|j\rangle,\ i =0,1,2,..., N-1$ is nonzero, so for sufficiently small $\varepsilon$, the number $\nu$ will also be nonzero, and for sufficiently small $\epsilon$, this number has the order $1/\varepsilon$.
\bigskip

We denote by $Z_{i, j}$ the set of occurrences of the letter $\varepsilon$ in the right part of the expression \eqref{hamapp}, and let $Z_j=\bigcup_iZ_{i, j}$. Then the number of elements in the set $Z_j$ will be $\nu$. 

We consider a smaller value of the quantum: $\epsilon=\varepsilon/\nu$. Substitute in the expression \eqref{quanta_expansion} instead of each occurrence of $\varepsilon$ its formal decomposition of the form $\varepsilon=\overbrace{\epsilon+\epsilon+\ldots +\epsilon}^\nu$, obtaining the decomposition of the amplitudes of the initial state on the smaller summands:

\begin{equation}
\label{refined_expansion}
\begin{array}{ll}
\la_j=&\langle j|\Psi\rangle\approx sign_{re}( \underbrace{\overbrace{\epsilon+\epsilon+\ldots +\epsilon}^\nu+\overbrace{\epsilon+\epsilon+\ldots +\epsilon}^\nu+\ldots +\overbrace{\epsilon+\epsilon+\ldots +\epsilon}^\nu}_{M_j})+\\
&sign_{im}i( \underbrace{\overbrace{\epsilon+\epsilon+\ldots +\epsilon}^\nu+\overbrace{\epsilon+\epsilon+\ldots +\epsilon}^\nu+\ldots +\overbrace{\epsilon+\epsilon+\ldots +\epsilon}^\nu}_{N_j} )
\end{array}
\end{equation}

Let $W^j_1, W^j_2,..., W^j_{m_j+N_j}$ be the sets of occurrences of the letter $\epsilon$ to the right side of the expression \eqref{refined_expansion}, marked with upper braces. Each of these sets has $\nu$ elements, as in the previously defined sets $Z_j$. Therefore, we can construct for each such set $W^j_s$ a one-to-one mapping of the form $\xi:\ W_s^j\rightarrow Z_j$. For each occurrence of $\varepsilon$ in \eqref{quanta_expansion}, its descendants are naturally defined: there are occurrences of $\epsilon$ in \eqref{refined_expansion}; there are $\nu$ descendants for each such occurrence.

For each pair of the form $(w_s^j,\xi (w_s^j))$, where $w_s^j\in W_s^j$, we will match the state transition and the type transition naturally. Namely, the transition of states will have the form $j\rightarrow i$ for such $i$ that $\xi (w_s^j)\in Z_{i,j}$; the transition of types $t_{in}\rightarrow t_{fin}$ is defined so that $t_{in}$ is the type of occurrence\footnote{the type of occurrence is determined naturally after opening the brackets, for example, for occurrence $...- i\epsilon ...$ type will be $ - i$.} $w_s^j$, and type $t_{fin}$ is the product of the type of occurrence $t_{in}$ and the type of occurrence $\xi(w_s^j)$. Sets $W^j_s$ do not intersect for different pairs $j,s$, so we set the domain of the function $\xi$ all occurrences of the letter $\epsilon$ in the right part of \eqref{refined_expansion} (see Fig. 1).

We will map each occurrence $\epsilon$ in the expression \eqref{refined_expansion} unique identifier and determine its corresponding quantum so that a) the initial state and the initial type of this quantum correspond to this occurrence and b) the transition of states and types for this quantum corresponds to the mapping $\xi$ in the above sense. Condition {\bf Q} will be satisfied, as in the expression for the matrix element of \eqref{hamapp} there are no cancelling members. Therefore, we have determined the amplitude quantization.

Then point 1 of Theorem will be satisfied by the initial choice of partition \eqref{quanta_expansion}. By virtue of our definition of the function $\xi$, the distribution of amplitudes in the state $|\theta\Psi\rangle$ will be proportional to the distribution of amplitudes in the state $cH|\Psi\rangle$ for any constant $c>0$. In fact, we are talking about choosing the time value $t=c$ in the action of the operator $tH$ on the initial state. In order to determine the value of $c$, which is reguired in item 2, we calculate the deposit of each occurrence of $\varepsilon^2$ to the right side of the equality \eqref{mult} and compare it with the deposit of the corresponding letter $\epsilon$ in $ |\theta\Psi\rangle$.

We fix some type transition $t_{in}\rightarrow t_{fin}$ and state transition $s_{in}\rightarrow s_{fin}$. We call the occurrence of $\varepsilon^2$ in the result of disclosure of brackets in \eqref{mult} corresponding to these transitions, if $j=s_{in},\ i=s_{fin}$, and this occurrence is obtained by multiplying the occurrence of $\varepsilon$ of type $t_{in}$ in the first factor of the right side of \eqref{mult} with the occurrence of $\varepsilon$ in the second factor of type $t'$, so $t_{in}t'=t_{fin}$. Each such occurrence $\varepsilon^2 $ corresponds to exactly one quantum amplitude of size $ \epsilon $ from the quantization of the amplitude defined above through the function $ \xi$, which has the same transitions of states and types: this quantum corresponds to the occurrence$ \epsilon $ which by one-to-one mapping $ \xi $ transforms into this occurrence $\varepsilon^2$. Therefore, the target value $c$ can be obtained from the ratio $\varepsilon^2/1=\epsilon/ c$, where, due to $\epsilon=\varepsilon / \nu$, we obtain $c=1 / \nu\varepsilon$, which has the order 1.

Since the accuracy of the approximation of the final state by $\theta$- shift  in the order of magnitude coincides with $\varepsilon$, we obtain inequality \eqref{passage_expansion}. Theorem is proved.

\begin{figure}
\begin{center}
\includegraphics[height=0.5\textwidth]{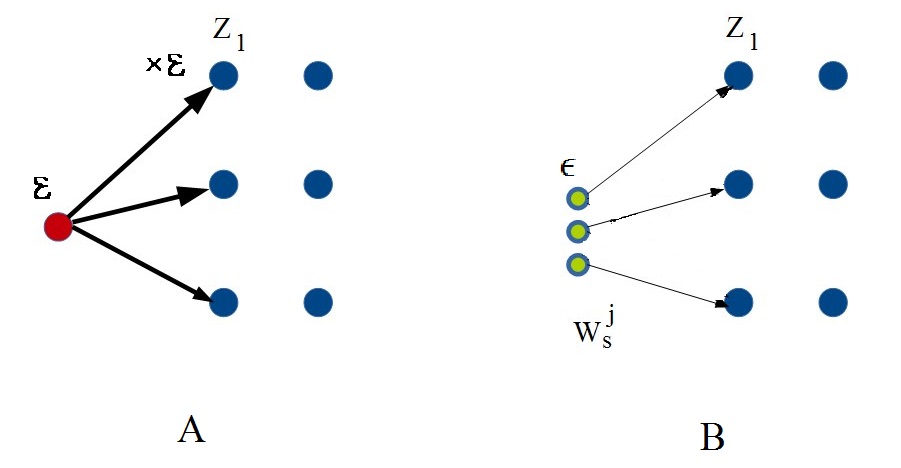}
\caption{A. Multiplying the state vector by a matrix. The contribution of each occurrence of $\varepsilon$ is multiplied by $\varepsilon$.
B. $\theta$- shift of the initial state. The size of the amplitude quantum $\epsilon$ has the order $\varepsilon^2$. }
\label{fig:nps}
\end{center}
\end{figure}

The following Corollary directly follows from Theorem.
\bigskip

{\it Corollary

In the conditions of Theorem  $\|\ |\theta\Psi\rangle-c H|\Psi\rangle\|$ has the order $\varepsilon$.  }
\bigskip

The corollary means that we can attribute to each amplitude quantum its unique history, that is, match it with a portion of the amplitude in the state $c H |\Psi\rangle$, which is in the natural sense a descendant of a given quantum. In particular, we can say that two quantum amplitudes cancel at $\theta$- shift, if their descendants cancel.

\section{Quantization of the amplitudes for heterogeneous ensembles}

To what extent can the result on homogeneous ensembles be generalized? We consider as an example again the Tavis-Cummings models with RWA approximation, and let now $|\Psi\rangle=\sum\limits_j\la_j | j\rangle$ be an arbitrary state of atoms and fields in the optical cavity. We have seen that the homogeneity of the ensemble entails the same flow of time-a single parameter $\tau$ for the whole ensemble. For ensembles of atoms with the different force of interaction with the field we have Example 1 (see above) of connected states, in which time actually flows uniformly for all particles. In general, we will no longer have this convenience, and to generalize the micro-causality we will have to redefine the amplitude quanta anew for small periods of time.

The heterogeneity of dynamics has two sources: the different forces of interaction of different atoms with the field - the heterogeneity of ensemble, and the different parts of the wave function - the heterogeneity of the initial state. Let $Q\subseteq \{ 1,2,...,n\}$ be such a subset of all qubits that the force of their interaction with the field is the same - we denote it by $g(Q)$, and $Q$ cannot be extended with the conservation of this property. Let $v$ be the set of values for qubits not belonging to $Q$, $J_{Q,v}$ is such a set of basic states, in which the values of qubits not belonging to $Q$, coincide with $v$. 

The pair $Q,v$ corresponds to the part of Hamiltonian, which depends only on $Q$; we denote it by  $H_{Q}$, it contains those and only those summands $\hbar\w\s_j^+\s_j$, for which $j\in Q$, there is no energy of the field and it contains exactly those summands of interaction $g(Q)(\s_j^+a+\s_ja^+)$, for which $j\in Q$.

The Hamiltonian then can be represented as the sum  
\begin{equation}
\label{hamsplit}
H=H_0+\sum\limits_{Q}H_{Q}, 
\end{equation}
where $H_0=\hbar a^+a$ and as usual, it is assumed that on the missing variables the summands of operators act as identical, that is, for example, under $\hbar a^+a$ is understood $\hbar a^+a\otimes I_{at}$ ($i_{at}$ - identical operator on atomic qubits).

The state $|\Psi\rangle$ we can represent as the sum of orthogonal summands of the form  \begin{equation}
\label{statesplit}
|\Psi\rangle=\sum\limits_{Q,v}\la_{Q,v}|\phi_{Q,v}\rangle,
\end{equation}
where $|\phi_{Q,v}\rangle\in{\cal H}_{Q,v}$, and the expansion \eqref{statesplit} is not uniquely determined in contrast to \eqref{hamsplit}, which is determined uniquely. 

Now the result of the action of the original Hamiltonian $H$ on the state $ |\Psi\rangle$ can be represented as a linear combination of its actions on each component of the decomposition \eqref{statesplit}.  In turn, the action of the unitary evolution operator $e^{-iHt / \hbar}$ on this component $|\phi_{Q_0,v}\rangle$ by virtue of the Trotter formula can be approximately represented as
\begin{equation}
H|\phi_{Q_0,v}\rangle\approx (e^{-iH_0\d t/\hbar}e^{-iH_{Q_0}\d t/\hbar}e^{-iH_{Q_1}\d t/\hbar}...e^{-iH_{Q_L}\d t/\hbar})^{t/\d t}
\label{genexp}
\end{equation}
where $Q_0,Q_1,...,Q_L$- possible choises of subsets of $Q$, $\d t\ll t$ is a small value. Time thus turns out to be divided into small sections of length $\d t$, on each of which only one component of Hamiltonian actually acts on the current state, and it affects either the set of atomic qubits, identical in relation to $H$, that is, their permutations commute with $H$ - in the case of $Q_0$, or, in the case of $Q_i,\ i>0$ - only one atom. In any case, on such a small interval of duration $\d t$ there will be micro-causality.

We see that for inhomogeneous ensembles, the introduction of micro-causality depends both on the atoms themselves and on the components of the initial state. Thus, the unconditional micro-causality takes place only in homogeneous ensembles.

\section{Conclusion}

We have established the presence of micro-causality for homogeneous ensembles of two-level particles interacting with the field. The necessity of homogeneity of the considered ensemble follows from the different nature of the flow of time for different parts of this ensemble, which makes it impossible to uniformly quantize the amplitude of ensembles consisting of differentiable particles.

Thus, scaling of any theory in which we want to preserve a micro - causality should be based on the same type, equal particles; only then the micro-causality remains with the increasing number of particles.

Such scaling to a complex system would allow a quantum way of describing the microcosm with rigid genetic determinism to be naturally coordinated in the description of very complex objects of biological nature, for the study of which such coordination is absolutely necessary.

\section{Acknowledgements}

The work is supported by the Russian Foundation for Basic Research, grant a-18-01-00695.

\end{document}